\PassOptionsToPackage{unicode}{hyperref}
\PassOptionsToPackage{hyphens}{url}
\documentclass[
]{article}
\usepackage{amsmath,amssymb}
\usepackage{iftex}
\ifPDFTeX
  \usepackage[T1]{fontenc}
  \usepackage[utf8]{inputenc}
  \usepackage{textcomp} 
\else 
  \usepackage{unicode-math} 
  \defaultfontfeatures{Scale=MatchLowercase}
  \defaultfontfeatures[\rmfamily]{Ligatures=TeX,Scale=1}
\fi
\usepackage{lmodern}
\ifPDFTeX\else
\fi
\IfFileExists{upquote.sty}{\usepackage{upquote}}{}
\IfFileExists{microtype.sty}{
  \usepackage[]{microtype}
  \UseMicrotypeSet[protrusion]{basicmath} 
}{}
\makeatletter
\@ifundefined{KOMAClassName}{
  \IfFileExists{parskip.sty}{%
    \usepackage{parskip}
  }{
    \setlength{\parindent}{0pt}
    \setlength{\parskip}{6pt plus 2pt minus 1pt}}
}{
  \KOMAoptions{parskip=half}}
\makeatother
\usepackage{xcolor}
\usepackage{longtable,booktabs,array}
\usepackage{multirow}
\usepackage{calc} 
\usepackage{etoolbox}
\makeatletter
\patchcmd\longtable{\par}{\if@noskipsec\mbox{}\fi\par}{}{}
\makeatother
\IfFileExists{footnotehyper.sty}{\usepackage{footnotehyper}}{\usepackage{footnote}}
\makesavenoteenv{longtable}
\usepackage{graphicx}
\makeatletter
\def\maxwidth{\ifdim\Gin@nat@width>\linewidth\linewidth\else\Gin@nat@width\fi}
\def\maxheight{\ifdim\Gin@nat@height>\textheight\textheight\else\Gin@nat@height\fi}
\makeatother
\setkeys{Gin}{width=\maxwidth,height=\maxheight,keepaspectratio}
\makeatletter
\def\fps@figure{htbp}
\makeatother
\ifLuaTeX
  \usepackage{selnolig}  
\fi
\usepackage{bookmark}
\IfFileExists{xurl.sty}{\usepackage{xurl}}{} 
\hypersetup{
  hidelinks,
  pdfcreator={LaTeX via pandoc}}

\author{}
\date{}

\usepackage{anyfontsize}
\usepackage[font= scriptsize, labelfont=bf]{caption}
\begin{document}

\begin{center}
\textbf{\fontsize{18}{24}\selectfont Motor Imagery Task Alters Dynamics of Human Body Posture}

\vspace{\baselineskip}
\textbf{\fontsize{10}{24}\selectfont Fatemeh Delavari, Seyyed Mohammad Reza Hashemi
Golpayegani, Mohammad Ali
Ahmadi-Pajouh}

Biomedical Engineering Department, Amirkabir
University of Technology, Tehran, Iran

Fatemeh.delavari@uconn.edu,\{m.r.hashemig, pajouh\}@aut.ac.ir




\end{center}

\vspace{\baselineskip}
\emph{\textbf{Abstract}}

Motor Imagery (MI) is gaining traction in both rehabilitation and sports
settings, but its immediate influence on human postural control is not
yet clearly understood. The focus of this study is to examine the
effects of MI on the dynamics of the Center of Pressure (COP), a crucial
metric for evaluating postural stability.

In the experiment, thirty healthy young adults participated in four
different scenarios: normal standing with both open and closed eyes, and
kinesthetic motor imagery focused on mediolateral (ML) and
anteroposterior (AP) sway movements. A mathematical model was developed
to characterize the nonlinear dynamics of the COP and to assess the
impact of MI on these dynamics.

Our results show a statistically significant increase
(p-value\textless0.05) in variables such as COP path length (PL) and
Long-Range Correlation (LRC) during MI compared to the closed-eye and
normal standing conditions. These observations align well with
psycho-neuromuscular theory, which suggests that imagining a specific
movement activates neural pathways, consequently affecting postural
control.

This study presents compelling evidence that motor imagery not only has
a quantifiable impact on COP dynamics but also that changes in the
Center of Pressure (COP) are directionally consistent with the imagined
movements. This finding holds significant implications for the field of
rehabilitation science, suggesting that motor imagery could be
strategically utilized to induce targeted postural adjustments.
Nonetheless, additional research is required to fully understand the
complex mechanisms that underlie this relationship and to corroborate
these results across a more diverse set of populations.


\vspace{\baselineskip}
\section{\texorpdfstring{\textbf{Introduction}}{Introduction}}\label{introduction}

Postural control, integral to most daily activities, is a multifaceted
task dependent on the synergy of various systems, notably the visual and
proprioceptive systems (1). Motor imagery, defined as the mental
reconstruction of movement without its actual execution (2), can be
categorized into two strategies: kinesthetic and visual. In kinesthetic
motor imagery (MI), individuals visualize themselves executing the
movement (first person perspective). In contrast, visual MI involves
picturing another person or oneself performing the action from a
third-person viewpoint (3). Research has identified overlapping
mechanisms between motor imagery and actual movement execution,
particularly in the realm of kinesthetic motor imagery (4). Moreover, MI
has shown promise in rehabilitation and sports domains due to its
benefits (5-12). Motor imagery is grounded in two primary theories: the
peripheral (often called psycho-neuromuscular) and the central theories.
The peripheral theory posits that imagining a specific movement results
in sub-threshold muscle activations related to that movement,
effectively simulating the neural pathways activated during actual
movement (13-15). Conversely, other research underscores the enhancement
in motor performance post MI sessions without noticeable muscle
activity, accentuating the shared brain regions\textquotesingle{}
activation during both imagined and actual movements (central theory).
This central viewpoint indicates overlapping neural pathways for both MI
and execution, especially within areas like the premotor and primary
motor cortex (16).

The relationship between motor imagery and postural sway has been the
focus of several studies (17-22). Some have posited that MI can induce
observable changes in postural control and sway. Yet, the underlying
mechanisms remain elusive. Is it cognitive distraction, neural resource
allocation, inhibitory circuits, the dual mechanism, or another factor
responsible for the changes in postural sway during MI? This study
endeavors to probe the relation between the direction of actual postural
sway and the imagined one. There are four possible hypotheses: 1) No
discernible change in postural sway occurs during MI and improvements in
postural control after MI sessions are attributed either to the
activation of brain regions common between MI and movement execution
(central theory) or to mentally rehearsing movement steps (symbolic
learning theory). 2) Imagining a movement stimulates identical motor
circuits as executing the movement, albeit to a lesser degree. Hence,
visualizing a specific sway direction might elicit a minimal yet
perceptible postural muscle response in that direction, aligning with
the psycho-neuromuscular theory. 3) If maintaining stable posture is
prioritized over the imagined sway, and/or if inhibitory circuits are
hyperactive during MI, the actual sway could be counteractive. 4)
Changes in postural sway might be observed, but these alterations might
not correspond to the imagined sway direction. Such a phenomenon could
stem from cognitive distractions or a lack of feedback mechanisms
governing postural sway during MI. Deciphering the relationship between
the direction of imagined and actual postural sway provides crucial
insights for rehabilitation. These insights are particularly valuable in
crafting therapy sessions that leverage MI and in monitoring the
progress of these sessions, especially when addressing issues of
compromised balance and postural control.

Balance assessment techniques span from rudimentary clinical tests to
advanced devices like force plates, electromyograms, and cameras. Force
plates, instrumental for biomechanical evaluations of gait and posture,
record the center of pressure (COP) (23). Within the force
plate\textquotesingle s context, the COP is the singular point where the
pressure field\textquotesingle s cumulative sum acts. This study employs
COP\textquotesingle s movement and positioning as a postural control
indicator.

This research delves into the effects of motor imagery on COP dynamics
by studying thirty healthy individuals during upright standing. Further,
we suggest a COP dynamics model to illustrate these effects. Although
numerous models exist for human posture (24-31), many view the human
body as a linear system---like an inverted pendulum---with attributes
such as time delay, feedback, and a proportional derivative controller.
However, Khanian et. al \textquotesingle s model (32) accounts for the
nonlinear dynamics of COP and can better explain and quantify the COP
changes. Consequently, we adapt this model in this study to illustrate
postural control during regular standing and MI and discuss its
potential implications for future occupational therapy and physiotherapy
sessions.

\section{\texorpdfstring{\textbf{Methods}}{Methods}}\label{methods}

\begin{enumerate}
\def\labelenumi{\arabic{enumi}.}
\item
  \textbf{Participants}
\end{enumerate}

Thirty healthy young adults, comprising twenty females and ten males,
participated in the study. These individuals had no history of
neurological, musculoskeletal, or vestibular disorders. They refrained
from taking any medications 24 hours prior to the experiment, though the
use of vitamins and dietary supplements was permissible. Participation
was voluntary, and all participants provided informed consent. They were
instructed to self-report their levels of mental and physical fatigue,
as well as their attention, before the experiment and after each testing
block. After each block, they also reported any discomfort experienced.
Due to excessive fatigue or diminished attention, six participants were
excluded from the analysis. The characteristics of the remaining 24
participants, including age, height, and weight, can be found in Table
1.

\begin{longtable}[]{@{}
  >{\raggedright\arraybackslash}p{(\columnwidth - 6\tabcolsep) * \real{0.3972}}
  >{\raggedright\arraybackslash}p{(\columnwidth - 6\tabcolsep) * \real{0.2009}}
  >{\raggedright\arraybackslash}p{(\columnwidth - 6\tabcolsep) * \real{0.2009}}
  >{\raggedright\arraybackslash}p{(\columnwidth - 6\tabcolsep) * \real{0.2009}}@{}}
\toprule\noalign{}
\caption{ Range, mean and standard deviation of
participants\textquotesingle{} age, height, and weight.}\\
\midrule\noalign{}
\begin{minipage}[b]{\linewidth}\raggedright
\end{minipage} & \begin{minipage}[b]{\linewidth}\raggedright
\textbf{Age (year)}
\end{minipage} & \begin{minipage}[b]{\linewidth}\raggedright
\textbf{Height (cm)}
\end{minipage} & \begin{minipage}[b]{\linewidth}\raggedright
\textbf{Weight (kg)}
\end{minipage} \\
\midrule\noalign{}
\endhead
\bottomrule\noalign{}
\endlastfoot
\textbf{Mean ± SD} & 24.6 ± 3.5 & 166.1 ± 6.9 & 60.6 ±
10.9 \\
\textbf{Range} & 22-34 & 152-179 & 42-92 \\
\end{longtable}

\begin{enumerate}
\def\labelenumi{\arabic{enumi}.}
\setcounter{enumi}{1}
\vspace{\baselineskip}
\item
  \textbf{Experimental apparatus and procedure}
\end{enumerate}

We utilized the Zebris FDM-S force plate to capture the COP data at a
sampling frequency of 120 Hz. Each recording session lasted 60 seconds.
During the experiments, participants stood barefoot, positioning their
feet together and keeping their arms by their sides. They were
instructed to look straight ahead, maintain an erect posture for the
neck and shoulders, and refrain from speaking.

The experiment consisted of four blocks, with a total duration of 15
minutes, as shown in Figure 1. In both the initial and final blocks,
participants\textquotesingle{} postural control was assessed during
standard standing under two conditions: with eyes open and eyes closed,
each lasting one minute. For blocks 2 and 3, participants engaged in
motor imagery of mediolateral (ML) and anteroposterior (AP) sways around
the ankle joint. Within these blocks, a three-minute familiarization
session was integrated. During this period, participants first executed
the sway with their eyes open and then practiced the imagery of the sway
with eyes closed. If the participant needed more practice, the training
duration was extended by a minute. Participants were instructed to
envision swaying around their ankle joint at the maximum angle where
they felt comfortable. The ankle joint was selected due to its critical
role in both postural control and gait. To mitigate any order effects on
the outcomes, the sequence in which participants performed the eyes-open
and eyes-closed tasks, as well as the tasks in blocks 2 and 3, was
randomized. Participants were allotted a one-minute rest interval
between blocks. Motor imagery was conducted with eyes closed to
eliminate potential visual feedback that might influence visualization
and to emphasize the role of proprioception. To ascertain that the
effects noted in the MI blocks weren\textquotesingle t simply a result
of having the eyes closed, we compared postural control during motor
imagery to postural control in both eyes-closed and eyes-open
conditions.




\begin{figure}[t]
	\centerline{\includegraphics[width=\textwidth]{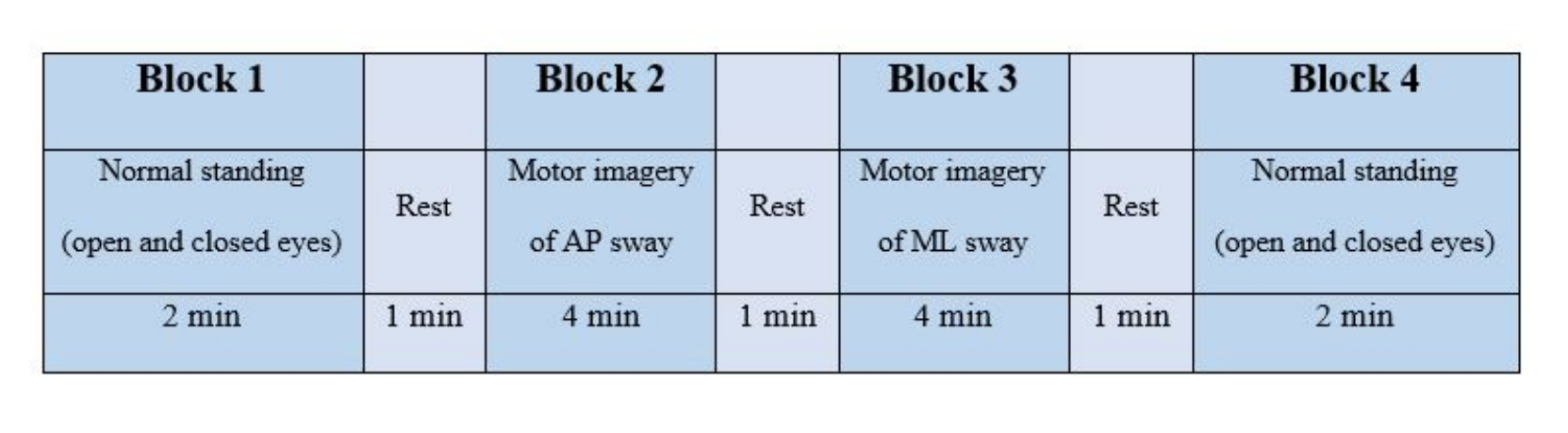}}
	\caption{An overview of the protocol. The experiment consisted of 4 blocks with a total duration of 15 minutes. In the first and final
blocks, postural control of normal standing was recorded in two
conditions of open and closed eyes, each for one minute. In the second
and third blocks, the participants were asked to have motor imagery of
sway around the ankle joint in ML and AP directions. In these two
blocks, three minutes of training were considered, in which participants
became familiar with the task by performing the sway with open eyes and
then practicing the imagination of the sway with closed eyes. The order
of performing the open and closed eyes tasks, as well as the blocks 2
and 3, were random for preventing the effects of precedence on the
results. One minute of rest was considered between the blocks.}
	\label{fig}
\end{figure}

\vspace{\baselineskip}
\vspace{\baselineskip}
\begin{enumerate}
\def\labelenumi{\arabic{enumi}.}
\setcounter{enumi}{2}
\item
  \textbf{Data analysis}
\end{enumerate}
The COP time series were analyzed in both the AP and ML directions,
using the features outlined below.

\begin{enumerate}
\def\labelenumi{\arabic{enumi})}
\item
  \emph{Path lengt (PL)}
\end{enumerate}

PL is one of the most common features for analyzing the changes
in the COP (33) and is calculated as:

\begin{longtable}[]{@{}
  >{\raggedright}p{(\columnwidth - 2\tabcolsep) * \real{0.8000}}
  >{\raggedright}p{(\columnwidth - 2\tabcolsep) * \real{0.2000}}@{}}
\begin{minipage}[b]{\linewidth}\raggedright
\[Path\ length = \sum_{i = 1}^{N}\sqrt{{(x_{i + 1} - x_{i})}^{2} + {(y_{i + 1} - y_{i})}^{2}}\]
\end{minipage} & \begin{minipage}[b]{\linewidth}\raggedleft
(1)
\end{minipage} \\
\end{longtable}


In equation 1, \emph{i} represents the sample number, and \emph{x} and
\emph{y} are the COP coordinates in the ML and AP directions,
respectively.
\begin{enumerate}
\def\labelenumi{\arabic{enumi})}
\setcounter{enumi}{1}
\item
  \emph{Correlation dimension (CD)}
\end{enumerate}

Earlier research indicates that the center of pressure (COP) in a
standing individual displays chaotic (34, 35). The correlation dimension
is a feature commonly used in chaotic signal analysis. It quantifies the
fractality of a signal by determining the average number of points
within a specific radial distance (R) from each point in the phase space
(36, 37). Through this method, we aim to delve into the nonlinear
dynamics and chaotic attributes of the COP.

\begin{longtable}[]{@{}
  >{\raggedright\arraybackslash}p{(\columnwidth - 2\tabcolsep) * \real{0.8}}
  >{\raggedright\arraybackslash}p{(\columnwidth - 2\tabcolsep) * \real{0.2}}@{}}\noalign{}
\begin{minipage}[b]{\linewidth}\raggedright
\[P_{i}(R) = \frac{1}{N - 1}\sum_{j = 1,j \neq i}^{N}{\Theta(R - \sqrt{{(x_{i} - x_{j})}^{2} + {(y_{i} - y_{j})}^{2}})}\]
\end{minipage} & \begin{minipage}[b]{\linewidth}\raggedleft
(2)
\end{minipage} \\
\(C(R) = \) 
& \begin{minipage}[b]{\linewidth}\raggedleft
(3)
\end{minipage} \\
\(\frac{1}{N(N - 1)}\sum_{i = 1}^{N}{\sum_{j = 1,j \neq i}^{N}{\Theta\left( R - \sqrt{{(x_{i} - x_{j})}^{2} + {(y_{i} - y_{j})}^{2}} \right)}}\)
\\
\(D_{c} = \lim_{R \rightarrow 0}\frac{\log{C(R)}}{\log R}\) 
& \begin{minipage}[b]{\linewidth}\raggedleft
(4)
\end{minipage} \\
\end{longtable}

In equations 2 and 3, \emph{x} and \emph{y} show the COP coordinates in the phase space. \emph{R} is a small radius, \emph{$\Theta$} is a Heaviside
function, and N denotes the total number of points. \(P_{i}(R)\) is the
possibility of points to lie in the given distance \emph{(R)} from the
i-th point. In equations 3 and 4, \emph{C(R)} is the mean of
\(P_{i}(R)\) and, finally, \(D_{c}\) is the correlation dimension.

\begin{enumerate}
\def\labelenumi{\arabic{enumi})}
\setcounter{enumi}{2}
\item
  \emph{Mean velocity (MV)}
\end{enumerate}

MV of COP shifts in each direction was determined by
averaging the displacement for each time interval, as follows:

\begin{longtable}[]{@{}
  >{\raggedright\arraybackslash}p{(\columnwidth - 2\tabcolsep) * \real{0.8000}}
  >{\raggedright\arraybackslash}p{(\columnwidth - 2\tabcolsep) * \real{0.2000}}@{}}
\noalign{}
\begin{minipage}[b]{\linewidth}\raggedright
\[Mean\ velocity = F \times mean\ \left| x_{i + 1} - x_{i} \right|\]
\end{minipage} & \begin{minipage}[b]{\linewidth}\raggedleft
(5)
\end{minipage} \\

\end{longtable}

In equation 5, \emph{x} indicates the COP in ML or AP direction and
\emph{i} represents the sample number. \emph{F} is the sampling
frequency, which was 120 Hz in this study.

\begin{enumerate}
\def\labelenumi{\arabic{enumi})}
\setcounter{enumi}{3}
\item
  \emph{Root mean square (RMS)}
\end{enumerate}

We calculated the RMS for the mediolateral and anteroposterior COP
signal.

\begin{longtable}[]{@{}
  >{\raggedright\arraybackslash}p{(\columnwidth - 2\tabcolsep) * \real{0.5000}}
  >{\raggedright\arraybackslash}p{(\columnwidth - 2\tabcolsep) * \real{0.5000}}@{}}
\noalign{}
\begin{minipage}[b]{\linewidth}\raggedright
\[RMS = \sqrt{\frac{x_{1}^{2} + x_{2}^{2} + \ldots + x_{n}^{2}}{n}}\]
\end{minipage} & \begin{minipage}[b]{\linewidth}\raggedleft
(6)
\end{minipage} \\
\end{longtable}

In equation 6, \emph{x} and \emph{n} indicate the COP in each of ML or
AP directions at each timepoint and the number of samples, respectively.

\begin{enumerate}
\def\labelenumi{\arabic{enumi})}
\setcounter{enumi}{4}
\item
  \emph{Range}
\end{enumerate}

Range was defined as the difference between the maximum and minimum
amount of the COP during the test period. This was computed
independently for both the mediolateral and anteroposterior COP signals.


\begin{longtable}[]{@{}
  >{\raggedright\arraybackslash}p{(\columnwidth - 2\tabcolsep) * \real{0.5000}}
  >{\raggedleft\arraybackslash}p{(\columnwidth - 2\tabcolsep) * \real{0.5000}}@{}}
\noalign{}
\begin{minipage}[b]{\linewidth}\raggedright
\[Range = \ x_{\max} - x_{\min}\]
\end{minipage} & 
\begin{minipage}[b]{\linewidth}\raggedleft
(7)
\end{minipage} \\
\end{longtable}

In equation 7, \emph{x} indicates the COP in either direction.

\begin{enumerate}
\def\labelenumi{\arabic{enumi})}
\setcounter{enumi}{5}
\item
  \emph{Long-range correlation (LRC)}
\end{enumerate}

Detrended Fluctuation Analysis (DFA) gives a measure of long-range
correlation in chaotic signals (38). In DFA, at first, the infinite
function \emph{X} is obtained from the \emph{x} time series. Then, the
resulting time series is divided into \emph{N} non-overlapping windows.
The \emph{Y} line is fitted on each window by the least squares method.
The detrended time series for each window (i.e.,\(\ \ F(n)\) in
equations 9 and 10) is calculated as the difference between the time
series values and the fitted line on each piece. This amount is calculated for different window lengths (i.e., \emph{t} in equations 8 and 9). Finally, the slope of the logarithmic diagram of the detrended time series versus the window length is considered as \emph{$\alpha$}, which indicates the degree of self-similarity.

\begin{longtable}[]{@{}
  >{\raggedright\arraybackslash}p{(\columnwidth - 2\tabcolsep) * \real{0.5000}}
  >{\raggedright\arraybackslash}p{(\columnwidth - 2\tabcolsep) * \real{0.5000}}@{}}\noalign{}
\begin{minipage}[b]{\linewidth}\raggedright
\[X_{t} = \sum_{i = 1}^{t}\left( x_{i} - mean(x) \right)\]
\end{minipage} 
& \begin{minipage}[b]{\linewidth}\raggedleft
(8)
\end{minipage} \\
\noalign{}
\endhead
\noalign{}
\endlastfoot
\(F(n) = \sqrt{\frac{1}{N}\sum_{t = 1}^{N}{(X_{t} - Y_{t})}^{2}}\) &
\begin{minipage}[b]{\linewidth}\raggedleft
(9)
\end{minipage} \\
\(F(n) \propto n^{\alpha}\) & 
\begin{minipage}[b]{\linewidth}\raggedleft
(10)
\end{minipage} \\
\end{longtable}

\vspace{\baselineskip}
\vspace{\baselineskip}
\vspace{\baselineskip}

\begin{enumerate}
\def\labelenumi{\arabic{enumi}.}
\setcounter{enumi}{3}
\vspace{\baselineskip}
\item
  \textbf{Statistical analysis}
\end{enumerate}

The statistical test was performed using repeated measures analysis of
variance (ANOVA). The Mauchly test was used to examine the sphericity of
the data, and in case of non-spherical data, the Greenhouse-Geisser
adjustment corrected the degrees of freedom. After examining the results
of the ANOVA test with the significance level of 0.05, which was
obtained in all cases, Tukey\textquotesingle s honest significant
difference performed post hoc comparisons.

\begin{enumerate}
\def\labelenumi{\arabic{enumi}.}
\setcounter{enumi}{4}
\item
\vspace{\baselineskip}
  \textbf{Model}

\end{enumerate}
\textbf{Model description}

We propose a COP dynamics model to illustrate postural control system
and the effects of motor imagery on this system. We adapt Khanian et. al
\textquotesingle s model for the human posture stability system (32)
which captures the nonlinear dynamics of COP and can properly explain
and quantify the COP changes. The proposed model is a two-dimensional
discrete-time chaotic map. The process equation which is a model for
creative processes (39) is used as the basis of the COP dynamics model
(equation 11).

\begin{longtable}[]{@{}
  >{\raggedright\arraybackslash}p{(\columnwidth - 2\tabcolsep) * \real{0.5000}}
  >{\raggedright\arraybackslash}p{(\columnwidth - 2\tabcolsep) * \real{0.5000}}@{}}
\noalign{}
\begin{minipage}[b]{\linewidth}\raggedright
\[x_{k + 1} = x_{k} + g \times \sin x_{k}\]
\end{minipage} & \begin{minipage}[b]{\linewidth}\raggedleft
(11)
\end{minipage} \\

\end{longtable}

This map has a biotic behavior for a wide range of the parameter
\emph{g} with a lot of periodic and instability windows. Bios is an
expansive pattern (40) and can model many structures, such as population
dynamics, air temperature and visual perception (41, 42). Since the
biotic pattern exists in the COP signal, this equation is appropriate
for modeling the COP signal in each of AP and ML directions. Considering
the interdependence of COP changes in both ML and AP directions, the
equations for these directions were coupled. As a result, equation 12
emerges as the model for COP dynamics.

\begin{longtable}[]{@{}
  >{\raggedright\arraybackslash}p{(\columnwidth - 2\tabcolsep) * \real{0.5000}}
  >{\raggedright\arraybackslash}p{(\columnwidth - 2\tabcolsep) * \real{0.5000}}@{}}
\noalign{}
\begin{minipage}[b]{\linewidth}\raggedright
\[x_{k + 1} = {a \times}e^{\frac{k}{q}} \times (x_{k} + b \times \sin y_{k})\]

\[y_{k + 1} = c \times e^{\frac{k}{q}} \times (y_{k} + d \times \sin x_{k})\]
\end{minipage} & \begin{minipage}[b]{\linewidth}\raggedleft
(12)
\end{minipage} \\

\end{longtable}

In this discrete-time map, \emph{x} and \emph{y} are the changes in the
COP signal in the ML and AP directions, respectively. Other parameters
such as a, b, c, d, and q are constants which should be identified using
optimization methods. The COP\textquotesingle s position at each time
step is influenced by its preceding location. Moreover, postural control
may decline due to physiological factors like fatigue and attention.
This results in an increased range of COP changes, which is accounted
for in this map through the exponential terms.

The model parameters are set as follows:

\begin{enumerate}
\def\labelenumi{\arabic{enumi})}
\item
  The bifurcation diagram was plotted for all parameters and the
  acceptable span for each of the parameters was determined. The range
  of COP changes in the recorded data was considered as a criterion for
  determining the acceptable span of parameters.
\item
  The genetic algorithm was used in order to set the parameters. In this
  method, the Euclidean distance was considered as the cost function.
  This method sets the parameters in a way that enhances the similarity
  between data and model phase spaces.
\end{enumerate}

\vspace{\baselineskip}
  \textbf{Effects of MI on COP dynamics}

During the imagery of front-back sway, parameter d exhibits a greater
increase than parameter b. Conversely, during lateral sway imagery,
parameter b rises more than parameter d. Indeed, the dependence of the
COP changes on the direction of the motor imagery is captured in the proposed model.

\vspace{\baselineskip}
  \textbf{Model validation}

To validate the model, we compared the range of COP changes between the
model and the data, as illustrated in Table 2. The range of COP changes
expands under closed eyes and motor imagery conditions. Further, when
imagining the ML and AP sways, the range of changes in the ML and AP
directions increases respectively. The model aptly captures this trend.
\setcounter{table}{1}


\begin{longtable}[]{@{}
  >{\raggedright\arraybackslash}p{(\columnwidth - 10\tabcolsep) * \real{0.2890}}
  >{\raggedright\arraybackslash}p{(\columnwidth - 10\tabcolsep) * \real{0.0735}}
  >{\raggedright\arraybackslash}p{(\columnwidth - 10\tabcolsep) * \real{0.1469}}
  >{\raggedright\arraybackslash}p{(\columnwidth - 10\tabcolsep) * \real{0.1616}}
  >{\raggedright\arraybackslash}p{(\columnwidth - 10\tabcolsep) * \real{0.1562}}
  >{\raggedright\arraybackslash}p{(\columnwidth - 10\tabcolsep) * \real{0.1727}}@{}}
  
 \toprule \noalign{}
\caption {A comparison between the range of changes in the model and the
data. For both the model and the data, range of changes increased in the
closed eyes and the motor imagery conditions. Plus, the range of changes
in ML and AP directions was higher while imagining the ML and AP sways,
respectively.
NS: normal standing with open eyes. CE: normal standing with closed
eyes. ML MI: mediolateral motor imagery. AP MI: anteroposterior motor imagery.}\\
\midrule\noalign{}
\begin{minipage}[b]
{\linewidth}\raggedright
\end{minipage} & \begin{minipage}[b]{\linewidth}\raggedright
\end{minipage} & \begin{minipage}[b]{\linewidth}\raggedright
\textbf{NS}
\end{minipage} & \begin{minipage}[b]{\linewidth}\raggedright
\textbf{CE}
\end{minipage} & \begin{minipage}[b]{\linewidth}\raggedright
\textbf{AP MI}
\end{minipage} & \begin{minipage}[b]{\linewidth}\raggedright
\textbf{ML MI}
\end{minipage} \\
\midrule\noalign{}
\endhead
\bottomrule\noalign{}
\endlastfoot
\multirow{2}{=}{\textbf{Range (Model)}

(Millimeter)} & \textbf{X} & 27.7 & 31.6 & 34.6 & 52.9 \\
& \textbf{Y} & 21.3 & 29.8 & 46.6 & 30.7 \\
\multirow{2}{=}{\textbf{Range (Data)}

(Millimeter)} & \textbf{X} & 26.8 & 33.45 & 35.2 & 54.8 \\
& \textbf{Y} & 24.7 & 30.5 & 42.6 & 37.6 \\
\end{longtable}

For the normal standing (NS) condition, we took the average of the range
of changes from NS1 (normal standing with open eyes during the first
block) and NS2 (normal standing with open eyes during the final block).
Similarly, for the closed eyes (CE) condition, we considered the average
range of changes between CE1 (normal standing with closed eyes in the
initial block) and CE2 (normal standing with closed eyes in the
concluding block).

Figure 2 presents a comparison of the COP phase space between the data
and the model. The data presented is the COP signal obtained from a
representative subject.


\begin{figure}[t]
	\centerline{\includegraphics[width=\textwidth]{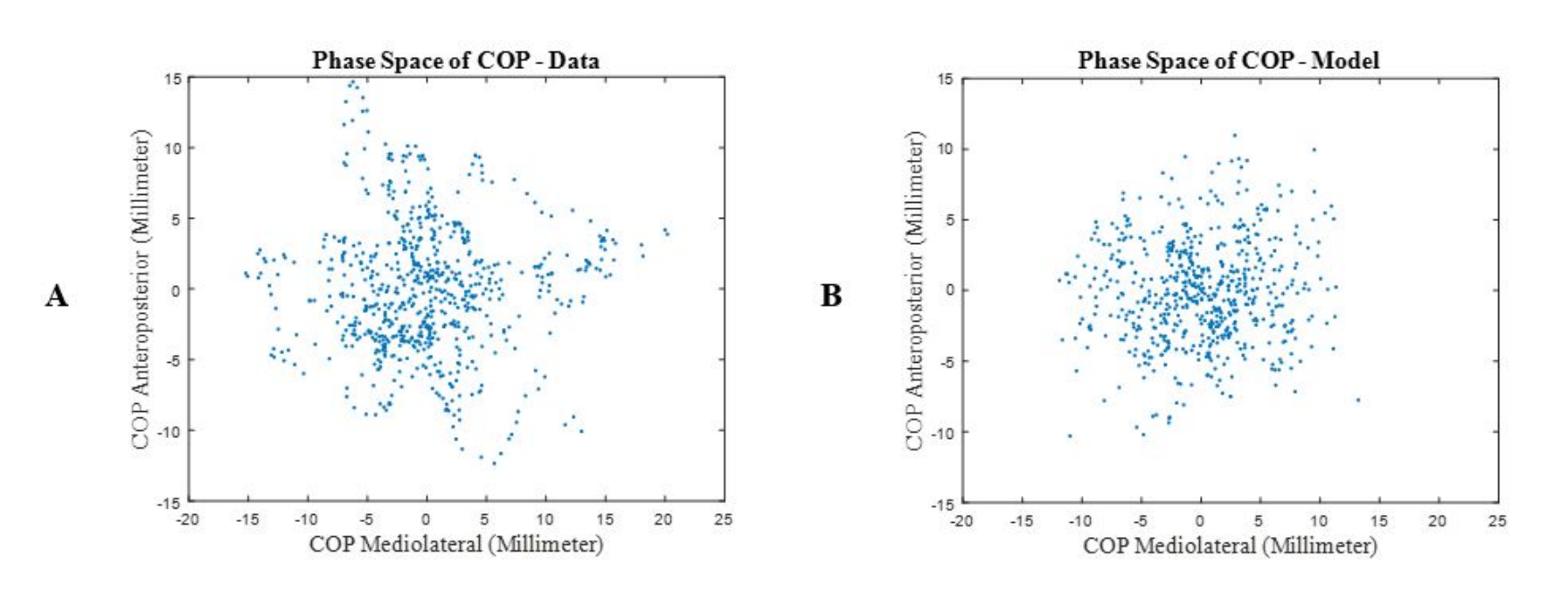}}
	\caption{A comparison between the phase space of COP for the data and
the model. A) Phase space of COP for the data of a representative
subject in the normal standing condition. B) Phase space of COP for the
model with parameters a=0.8, b=4.4, c=0.7, d=5.8 and q=4999.7.}
	\label{fig}
\end{figure}

\section{\texorpdfstring{\textbf{Results}}{Results}}\label{Results}



Relative to standing with closed eyes, the motor imagery condition
exhibited an increase in MV, range, RMS, LRC, and PL, all pointing to an augmented COP displacement. A similar trend
was observed when comparing standing with closed eyes to standing with
open eyes. In the CE condition, the absence of visual feedback leads to
greater COP displacement. Additionally, a decrease in the correlation
dimension was noted when comparing the MI condition to both the CE and
NS conditions, as well as when comparing CE to NS, indicating a
reduction in the system\textquotesingle s chaotic behavior. Typical
postural control exhibits a certain level of variability, reflecting the
body\textquotesingle s adaptability and responsiveness to slight
disturbances or environmental shifts. An overly regular postural control
(i.e., reduced chaos) may hint at a diminished capacity for adaptation.
Conversely, excessive chaotic behavior might imply instability. The
observed decrease in chaotic behavior during CE and MI conditions in the
results suggests a deviation from the normal postural control in these
scenarios. Table 3 provides detailed information on the mean and
standard deviation for each measure across all conditions.

\begin{longtable}[]{@{}
  >{\raggedright\arraybackslash}p{(\columnwidth - 14\tabcolsep) * \real{0.1638}}
  >{\raggedright\arraybackslash}p{(\columnwidth - 14\tabcolsep) * \real{0.0649}}
  >{\raggedright\arraybackslash}p{(\columnwidth - 14\tabcolsep) * \real{0.1139}}
  >{\raggedright\arraybackslash}p{(\columnwidth - 14\tabcolsep) * \real{0.1234}}
  >{\raggedright\arraybackslash}p{(\columnwidth - 14\tabcolsep) * \real{0.1330}}
  >{\raggedright\arraybackslash}p{(\columnwidth - 14\tabcolsep) * \real{0.1284}}
  >{\raggedright\arraybackslash}p{(\columnwidth - 14\tabcolsep) * \real{0.1316}}
  >{\raggedright\arraybackslash}p{(\columnwidth - 14\tabcolsep) * \real{0.1411}}@{}}
\caption{Mean of the measures in different
conditions. MV is reported in millimeters per sec. Range and Path length are reported in Millimeters. NS1: normal standing with open eyes in the first block. NS2:
normal standing with open eyes in the last block. CE1: normal standing
with closed eyes in the first block. CE2: normal standing with closed
eyes in the last block. ML MI: mediolateral motor imagery. AP MI:
anteroposterior motor imagery.}\\
\toprule\noalign{}
\begin{minipage}[b]{\linewidth}\raggedright
\end{minipage} & \begin{minipage}[b]{\linewidth}\raggedright
\end{minipage} & \begin{minipage}[b]{\linewidth}\raggedright
\textbf{NS1}
\end{minipage} & \begin{minipage}[b]{\linewidth}\raggedright
\textbf{CE1}
\end{minipage} & \begin{minipage}[b]{\linewidth}\raggedright
\textbf{APMI}
\end{minipage} & \begin{minipage}[b]{\linewidth}\raggedright
\textbf{MLMI}
\end{minipage} & \begin{minipage}[b]{\linewidth}\raggedright
\textbf{CE2}
\end{minipage} & \begin{minipage}[b]{\linewidth}\raggedright
\textbf{NS2}
\end{minipage} \\
\midrule\noalign{}
\endhead
\bottomrule\noalign{}
\endlastfoot
\multirow{2}{=}{\textbf{MV}
} & \textbf{X} & 58.4 & 84.8 & 78.8 & 139.3 & 81.7 & 57.7 \\
& \textbf{Y} & 48.1 & 70.3 & 104.9 & 83.4 & 70.5 & 48.8 \\
\multirow{2}{=}{\textbf{Range}
} & \textbf{X} & 27.2 & 32.1 & 35.2 & 54.9 & 35.7 & 26.4 \\
& \textbf{Y} & 23.5 & 29.2 & 42.6 & 37.6 & 32.0 & 26.0 \\
\multirow{2}{=}{\textbf{RMS}} & \textbf{X} & 4.8 & 5.8 & 6.9 & 10.8 & 6.8 & 5.2 \\
& \textbf{Y} & 4.4 & 5.2 & 8.8 & 7.5 & 5.9 & 5.4 \\
\textbf{PL}
 & \textbf{} & 492.1 & 716.1 & 857.7 & 1043.5 & 705.04 & 493.6 \\
\textbf{CD} & \textbf{} & 0.53 & 0.34 & 0.21 & 0.19 & 0.36 & 0.53 \\
\multirow{2}{=}{\textbf{LRC}} & \textbf{X} & 0.07 & 0.14 & 0.13 & 0.29 & 0.14 & 0.09 \\
& \textbf{Y} & 0.10 & 0.19 & 0.38 & 0.27 & 0.19 & 0.12 \\
\end{longtable}

MV in both the AP and ML directions, PL, correlation
dimension, and LRC values in both AP and ML directions consistently show
significant differences in the ML and CE scenarios compared to normal
standing (p-value \textless{} 0.05). In the MI condition, PL is
also consistently and significantly different from that in the CE
condition. Furthermore, LRC in the ML direction is significantly
different when performing ML MI, and not when performing AP MI, compared
to standing with closed eyes. Similarly, LRC in the AP direction during
AP MI shows a significant difference compared to standing with closed
eyes.


\section{\texorpdfstring{\textbf{Discussion}}{Discussion}}\label{Discussion}

The primary objective of our study was to scrutinize the immediate
effects of motor imagery (MI) on Center of Pressure (COP) dynamics
within a population of healthy adults. Employing the model initially
proposed by Khanian et al. (32), we formulated a specialized COP
dynamics model that showcased the nonlinear intricacies of postural
control during MI activities. Our results strongly validate the
psycho-neuromuscular theory, revealing notable changes in variables like
COP PL and Long-Range Correlation (LRC) during MI sessions as
compared to closed eyes (CE) and normal standing (NS) conditions. The
observed enhancements in COP displacement underline MI\textquotesingle s
substantial impact on postural control mechanisms.

Our findings have critical implications for rehabilitation, particularly
for therapies centered on enhancing balance and postural stability. The
model\textquotesingle s adaptability for capturing individual postural
dynamics holds promise for the development of patient-specific treatment
protocols. This could be particularly beneficial for patients with
compromised balance, such as stroke survivors, in optimizing MI-based
therapeutic strategies. Beyond rehabilitation, our model parameters
could function as viable biomarkers for not only gauging the efficacy of
MI-centered interventions but also for evaluating postural control among
varying populations, including athletes and the elderly.

Our findings indicate that alterations in the body\textquotesingle s
center of pressure (COP) align directionally with the imagined
movements. Thus, the observed increase in COP changes during motor
imagery is not attributable to mental load or the lack of a feedback
mechanism. This insight has important implications for rehabilitation
science, as it suggests that motor imagery can be strategically employed
to elicit specific postural adjustments.

The results of this study align well with previous research,
particularly the study conducted by Rodrigues et al. (17), who focused
on motor imagery of bilateral plantar flexions. Rodrigues and colleagues
observed that kinesthetic imagery significantly increased COP
oscillations. These findings are consistent with other studies that have
demonstrated marked changes in postural sway during MI tasks.
Specifically, there is a noticeable increase in the center of pressure
(COP) displacement when standing subjects engage in kinesthetic imagery,
imagining themselves rising on their tiptoes (43). The congruence
between our results and prior research further validates the influence
of motor imagery on postural control and opens avenues for its
application in rehabilitative settings.

While our findings offer substantial evidence supporting
MI\textquotesingle s influence on COP dynamics, they are not without
limitations. One major limitation is the demographic homogeneity of our
study group, comprising solely of healthy young adults. Future studies
could enrich these findings by involving a more diverse participant
pool, like the elderly or individuals with neurodegenerative disorders.
Additionally, our study did not thoroughly account for external factors
like fatigue and attention, which were only self-reported by
participants. Subsequent research should aim to objectively measure
these variables.

Looking ahead, it would be useful to reconcile the psycho-neuromuscular
and central theories by examining the neural foundations of observed COP
modifications. Longitudinal studies could further elucidate the
long-term benefits of MI training on postural control, thereby cementing its therapeutic potential.

To sum up, our study provides robust evidence underscoring the role of
motor imagery in affecting postural control, as demonstrated through
measurable changes in COP dynamics. Leveraging a validated model, we
have quantitatively corroborated these shifts, thus reinforcing
MI\textquotesingle s potential impact on enhancing postural stability.
These findings lay a crucial foundation for future research and offer a
substantive rationale for the integration of MI into rehabilitation
programs, particularly those aiming at improving balance and postural
control.

The ramifications of our work are especially noteworthy for the field of
rehabilitative medicine. Given MI\textquotesingle s non-invasive
character and its proven efficacy in modifying COP dynamics, it emerges
as a promising supplement to existing therapeutic strategies for
improving postural stability, especially among those grappling with
balance-related disorders or conditions.

However, while the study sheds new light on MI\textquotesingle s
influence over postural control, it also indicates the need for more
in-depth research to unpack the complex mechanisms governing this
relationship. Enhanced understanding of these processes could streamline
the optimization of MI-based treatments. Furthermore, broadening the
scope to include various demographics is imperative for verifying the
universal applicability and adaptability of MI-centered therapies.

In summary, our research marks a significant advancement in the
application of motor imagery for rehabilitative therapies focused on
balance and postural stability. It offers a robust framework, backed by
empirical data, which future studies can refine and extend to understand
the underlying mechanisms better and to assess MI\textquotesingle s
utility across diverse populations.

  \textbf{Implications on Physiotherapy Practice}

The study demonstrates that motor imagery (MI) significantly impacts the
dynamics of the Center of Pressure (COP), a key metric in evaluating
postural stability. This finding can be directly applied in
physiotherapy practices, especially in designing rehabilitation
protocols for patients with balance and postural control issues.
Physiotherapists can incorporate MI tasks into their treatment plans to
induce specific postural adjustments, thus enhancing overall stability
and balance in patients.

The research shows that MI can alter COP dynamics in a directionally
consistent manner with the imagined movements. This insight is
particularly beneficial for rehabilitation programs aimed at patients
suffering from balance disorders, such as those recovering from stroke
or neurological injuries. Tailoring MI exercises to the specific needs
of these patients can aid in quicker and more effective recovery of
postural control.

Given the non-invasive nature of MI, it emerges as a promising
adjunctive tool in physiotherapy. It can be used alongside conventional
physical therapies, especially in cases where physical movement might be
limited or painful. This makes it an excellent option for early-stage
rehabilitation or for patients with severe mobility restrictions.

The adaptability of the MI approach, as demonstrated by the
study\textquotesingle s model, suggests that physiotherapists can
develop individualized treatment protocols. These personalized plans can
be more effective as they can be designed to target the specific
postural and balance issues of each patient, considering their unique
COP dynamics.

The study's findings provide a quantitative basis for understanding the
changes in postural control due to MI. This can help physiotherapists
not only in planning treatment but also in monitoring progress. The
measurable changes in COP dynamics offer a way to objectively assess the
effectiveness of the MI-based interventions over time.

The study highlights the need for further research, particularly
involving diverse demographics and long-term effects. This opens up
avenues for physiotherapists to engage in or follow up on ongoing
research, keeping abreast of the latest developments in the field which
can then be translated into practice.

In conclusion, this study's findings on the effects of motor imagery on
postural control offer valuable insights and practical applications for
physiotherapy practice, particularly in enhancing balance and postural
stability in various patient populations.

\vspace{\baselineskip}
\textbf{Declaration of Generative AI and AI-assisted technologies in the
writing process}

During the preparation of this work the authors used ChatGPT in order to
improve the readability of the manuscript. After using this
tool/service, the authors reviewed and edited the content as needed and
take full responsibility for the content of the publication.

\vspace{\baselineskip}
\textbf{Acknowledgments}

The authors thank Dr. Nasser Fatouraee and Arman Gholizadeh from the
Biomechanics department of Amirkabir University of Technology, Tehran,
Iran

\vspace{\baselineskip}
\textbf{Conflict of interest}

None declared.

\section{\texorpdfstring{\textbf{References}}{References}}\label{References}

1. Massion J. Postural control system. Current opinion in neurobiology.
1994;4(6):877-87.

2. Dickstein R, Deutsch JE. Motor imagery in physical therapist
practice. Physical therapy. 2007;87(7):942-53.

3. Grangeon M, Guillot A, Collet C. Postural control during visual and
kinesthetic motor imagery. Applied psychophysiology and biofeedback.
2011;36:47-56.

4. Mulder T. Motor imagery and action observation: cognitive tools for
rehabilitation. Journal of neural transmission. 2007;114:1265-78.

5. de Souza NS, Martins ACG, da Silva Canuto K, Machado D, Teixeira S,
Orsini M, et al. Postural control modulation during motor imagery tasks:
A systematic review. International Archives of Medicine. 2015;8.

6. Hecker JE, Kaczor LM. Application of imagery theory to sport
psychology: Some preliminary findings. Journal of Sport and Exercise
Psychology. 1988;10(4):363-73.

7. Morris T, Spittle M, Watt AP. Imagery in sport: Human Kinetics; 2005.

8. Oh D-S, Choi J-D. The effect of motor imagery training for trunk
movements on trunk muscle control and proprioception in stroke patients.
Journal of physical therapy science. 2017;29(7):1224-8.

9. Cho H-y, Kim J-s, Lee G-C. Effects of motor imagery training on
balance and gait abilities in post-stroke patients: a randomized
controlled trial. Clinical rehabilitation. 2013;27(8):675-80.

10. Fujikawa S, Ohsumi C, Ushio R, Tamura K, Sawai S, Yamamoto R, Nakano
H. Potential Applications of Motor Imagery for Improving Standing
Posture Balance in Rehabilitation. Neurorehabilitation and Physical
Therapy: IntechOpen; 2022.

11. Sarasso E, Agosta F, Piramide N, Gardoni A, Canu E, Leocadi M, et
al. Action observation and motor imagery improve dual task in
Parkinson\textquotesingle s disease: a clinical/fMRI study. Movement
Disorders. 2021;36(11):2569-82.

12. Chiacchiero M, Cagliostro P, DeGenaro J, Giannina C, Rabinovich Y.
Motor imagery improves balance in older adults. Topics in Geriatric
Rehabilitation. 2015;31(2):159-63.

13. Borah B, Yadav A. Effect of Psychoneuromuscular Theory and Visualization Technique in Reducing Anxiety Levels of Female Soccer Players in Competition Situations. International Journal of Behavioral
Social and Movement Sciences. 2017;6(2):1-6.

14. Livesay JR, Samaras MR. Covert neuromuscular activity of the
dominant forearm during visualization of a motor task. Perceptual and
motor skills. 1998;86(2):371-4.

15. Bakker FC, Boschker MSJ, Chung T. Changes in muscular activity while
imagining weight lifting using stimulus or response propositions.
Journal of Sport and Exercise Psychology. 1996;18(3):313-24.

16. Krüger B, Hettwer M, Zabicki A, de Haas B, Munzert J, Zentgraf K.
Practice modality of motor sequences impacts the neural signature of
motor imagery. Scientific reports. 2020;10(1):19176-.

17. Rodrigues EC, Lemos T, Gouvea B, Volchan E, Imbiriba LA, Vargas CD.
Kinesthetic motor imagery modulates body sway. Neuroscience.
2010;169(2):743-50.

18. Stins JF, Schneider IK, Koole SL, Beek PJ. The influence of motor
imagery on postural sway: differential effects of type of body movement
and person perspective. Advances in Cognitive Psychology.
2015;11(3):77-.

19. Stins JF, Michielsen ME, Roerdink M, Beek PJ. Sway regularity
reflects attentional involvement in postural control: Effects of
expertise, vision and cognition. Gait \& posture. 2009;30(1):106-9.

20. Mouthon A, Ruffieux J, Mouthon M, Hoogewoud HM, Annoni JM, Taube W.
Age-related differences in cortical and subcortical activities during
observation and motor imagery of dynamic postural tasks: an fMRI study.
Neural plasticity. 2018;2018.

21. Jahn K, Deutschländer A, Stephan T, Strupp M, Wiesmann M, Brandt T.
Brain activation patterns during imagined stance and locomotion in
functional magnetic resonance imaging. Neuroimage. 2004;22(4):1722-31.

22. Taube W, Mouthon M, Leukel C, Hoogewoud H-M, Annoni J-M, Keller M.
Brain activity during observation and motor imagery of different balance
tasks: an fMRI study. cortex. 2015;64:102-14.

23. Alvarenga R, Porto F, Braga R, Cantreva R, Espinosa G, Itaborahy A,
et al., editors. Construction and calibration of a low-cost force plate
for human balance evaluation. ISBS-Conference Proceedings Archive; 2011.

24. Pasma JH, Boonstra TA, Van Kordelaar J, Spyropoulou VV, Schouten AC.
A sensitivity analysis of an inverted pendulum balance control model.
Frontiers in computational neuroscience. 2017;11:99-.

25. Maurer C, Peterka RJ. A new interpretation of spontaneous sway
measures based on a simple model of human postural control. Journal of
neurophysiology. 2005;93(1):189-200.

26. Kot A, Nawrocka A, editors. Modeling of human balance as an inverted
pendulum. Proceedings of the 2014 15th International Carpathian Control
Conference (ICCC); 2014: IEEE.

27. Mergner T, Maurer C, Peterka RJ. A multisensory posture control
model of human upright stance. Progress in brain research.
2003;142:189-201.

28. Milton J, Cabrera JL, Ohira T, Tajima S, Tonosaki Y, Eurich CW,
Campbell SA. The time-delayed inverted pendulum: implications for human
balance control. Chaos: An Interdisciplinary Journal of Nonlinear
Science. 2009;19(2).

29. Asai Y, Tasaka Y, Nomura K, Nomura T, Casadio M, Morasso P. A model
of postural control in quiet standing: robust compensation of
delay-induced instability using intermittent activation of feedback
control. PLoS One. 2009;4(7):e6169-e.

30. Hilts WW, Szczecinski NS, Quinn RD, Hunt AJ, editors. Simulation of
human balance control using an inverted pendulum model. Biomimetic and
Biohybrid Systems: 6th International Conference, Living Machines 2017,
Stanford, CA, USA, July 26--28, 2017, Proceedings 6; 2017: Springer.

31. Suzuki Y, Nomura T, Casadio M, Morasso P. Intermittent control with
ankle, hip, and mixed strategies during quiet standing: a theoretical
proposal based on a double inverted pendulum model. Journal of
theoretical biology. 2012;310:55-79.

32. Khanian MYA, Golpayegni SMRH, Rostami M. A new multi-attractor model
for the human posture stability system aimed to follow self-organized
dynamics. Biocybernetics and Biomedical Engineering. 2020;40(1):162-72.

33. Donker SF, Roerdink M, Greven AJ, Beek PJ. Regularity of
center-of-pressure trajectories depends on the amount of attention
invested in postural control. Experimental brain research.
2007;181:1-11.

34. Yamada N. Chaotic swaying of the upright posture. Human movement
science. 1995;14(6):711-26.

35. Pascolo PB, Marini A, Carniel R, Barazza F. Posture as a chaotic
system and an application to the Parkinson's disease. Chaos, Solitons \&
Fractals. 2005;24(5):1343-6.

36. Cimolin V, Galli M, Rigoldi C, Grugni G, Vismara L, Mainardi L,
Capodaglio P. Fractal dimension approach in postural control of subjects
with Prader-Willi Syndrome. Journal of neuroengineering and
rehabilitation. 2011;8:1-6.

37. Nygård JF, Glattre E. Fractal analysis of time series in
epidemiology: Is there information hidden in the noise? Norsk
Epidemiologi. 2003;13(2).

38. González-Salas JS, Shbat MS, Ordaz-Salazar FC, Simón J. Analyzing
chaos systems and fine spectrum sensing using detrended fluctuation
analysis algorithm. Mathematical Problems in Engineering. 2016;2016.

39. Sabelli H. Bios: a study of creation (With CD-ROM): World
Scientific; 2005.

40. Sabelli H, Kovacevic L. Biotic complexity of population dynamics.
Complexity. 2008;13(4):47-55.

41. Patel M, Sabelli H. Autocorrelation and frequency analysis
differentiate cardiac and economic bios from 1/f noise. Kybernetes.
2003;32(5/6):692-702.

42. Beigzadeh M, Golpayegani SMRH, editors. A macroscopic chaotic model
of visual perception. 2016 23rd Iranian Conference on Biomedical
Engineering and 2016 1st International Iranian Conference on Biomedical
Engineering (ICBME); 2016: IEEE.

43. Lemos T, Rodrigues EC, Vargas CD. Motor imagery modulation of
postural sway is accompanied by changes in the EMG--COP association.
Neuroscience Letters. 2014;577:101-5.

\end{document}